\documentclass[aps,prd,preprint,superscriptaddress,showpacs,floatfix,nofootinbib]{revtex4}

\usepackage{graphicx}
\usepackage{amsmath}
\usepackage{amssymb}

\newcommand{\Z}{{\mathbb Z}}

\newcommand{\bmat}{\left(\begin{array}}
\newcommand{\emat}{\end{array}\right)}
\newcommand{\beq}{\begin{equation}}
\newcommand{\eeq}{\end{equation}}

\def\yzero{\smash{\hbox{$y\kern-4pt\raise1pt\hbox{${}^\circ$}$}}}

\def\-{\hphantom{-}}

\def\s2{\frac{1}{\sqrt2}}

\def\beq{\begin{equation}}
\def\eeq{\end{equation}}
\def\beqa{\begin{eqnarray}}
\def\eeqa{\end{eqnarray}}

\def\IF{\relax{\rm I\kern-.18em F}}
\def\II{\relax{\rm I\kern-.18em I}}
\def\IP{\relax{\rm I\kern-.18em P}}

\def\Dsl{\,\raise.15ex\hbox{/}\mkern-13.5mu D} 

\def\IC{\bf C}
\def\IZ{\bf Z}

\def\z2z2{$\IC^3/(\IZ_2\times\IZ_2)$}

\def\s{\sigma}

\def\z{\zeta}

\def\bo{{\raise-.3ex\hbox{\large$\Box$}}}               
\def\face{{\raise.2ex\hbox{$\displaystyle \bigodot$}\mskip-2.2mu \llap {$\ddot
        \smile$}}}                                      

\def\leftrightarrowfill{$\mathsurround=0pt \mathord\leftarrow \mkern-6mu
        \cleaders\hbox{$\mkern-2mu \mathord- \mkern-2mu$}\hfill
        \mkern-6mu \mathord\rightarrow$}       
\def\dvec#1{\vbox{\ialign{##\crcr
        \leftrightarrowfill\crcr\noalign{\kern-1pt\nointerlineskip}
        $\hfil\displaystyle{#1}\hfil$\crcr}}}           

\def\beq{\begin{equation}}
\def\eeq{\end{equation}}

\def\beqx{\begin{displaymath}}
\def\eeqx{\end{displaymath}}

\def\beqa{\begin{eqnarray}}
\def\eeqa{\end{eqnarray}}

\begin{document}

\title{ Type IIB Supersymmetric Flux Vacua}

\author{Ching-Ming Chen}

\affiliation{Institute for Theoretical Physics, Vienna University
of Technology, Wiedner Hauptstrasse 8-10, 1040 Vienna, Austria}

\author{Shan Hu}

\affiliation{George P. and Cynthia W. Mitchell Institute for
Fundamental Physics, Texas A$\&$M University, College Station, TX
77843, USA }

\author{Tianjun Li}

\affiliation{Key Laboratory of Frontiers in Theoretical Physics,
      Institute of Theoretical Physics, Chinese Academy of Sciences,
Beijing 100190, P. R. China }

\affiliation{George P. and Cynthia W. Mitchell Institute for
Fundamental Physics, Texas A$\&$M University, College Station, TX
77843, USA }

\author{Dimitri V. Nanopoulos}

\affiliation{George P. and Cynthia W. Mitchell Institute for
Fundamental Physics, Texas A$\&$M University, College Station, TX
77843, USA }

\affiliation{Astroparticle Physics Group, Houston Advanced
Research Center (HARC), Mitchell Campus, Woodlands, TX 77381, USA}

\affiliation{Academy of Athens, Division of Natural Sciences, 28
Panepistimiou Avenue, Athens 10679, Greece }

\date{\today}

\begin{abstract}

On the Type IIB toroidal $\mathbf{T^6}$ orientifolds with generic
flux compactifications, we conjecture that in generic
supsersymmetric Minkowski vacua, at least one of the flux
contributions to the seven-brane and D3-brane tadpoles is positive
if the moduli are stabilized properly, and then the tadpole
cancellation conditions can not be relaxed. To study the
supsersymmetric Minkowski flux vacua, we simplify the fluxes
reasonably and discuss the corresponding superpotential. We show
that we can not have simultaneously the positive real parts of all
the moduli and the negative/zero flux contributions to all the
seven-brane and D3-brane tadpoles. Therefore, we can not construct
realistic flux models with the relaxed tadpole cancellation
conditions. When studying the supsersymmetric AdS vacua, we obtain
flux models with the seven-brane and D3-brane tadpole cancellation
conditions relaxed elegantly, and we present a semi-realistic
Pati-Salam model as well as its particle spectrum. The lifting
from the AdS vacua to the Minkowski/dS vacua remains a great
challenge in flux model buildings on toroidal orientifolds.

\end{abstract}


\preprint{ACT-09-11, MIFPA-11-29, TUW-11-16}

\maketitle


\section{Introduction}

The great challenge in string phenomenology is to construct
realistic string models without additional chiral exotic particles
at low energy and with the moduli fields stabilized.  In
particular, the intersecting D-brane models on Type II
orientifolds~\cite{Polchinski:1995df}, where the chiral fermions
arise from the intersections of D-branes in the internal
space~\cite{Berkooz:1996km} and the T-dual description in terms of
magnetized D-branes~\cite{Bachas:1995ik}, have been very
interesting during the last decade~\cite{Blumenhagen:2005mu}.
Further employing the renormalization group equations in these
models, we may test them at the Large Hadron Collider (LHC).

Initially many non-supersymmetric three-family Standard-like
models and Grand Unified Theories (GUTs)  were constructed on Type
IIA orientifolds with intersecting
D6-branes~\cite{Blumenhagen:2000wh, Angelantonj:2000hi,
Ibanez:2001nd, Blumenhagen:2005mu}. However, these models
generically suffer uncancelled Neveu-Schwarz-Neveu-Schwarz (NSNS)
tadpoles and the gauge hierarchy problem. Later, semi-realistic
supersymmetric Standard-like and GUT models were constructed in
Type IIA theory on the $\mathbf{T^6/(\Z_2\times \Z_2)}$
orientifold~\cite{CSU1, CSU2, CP, Cvetic:2002pj, CLL,
Cvetic:2004nk, Chen:2005ab, Chen:2005mj, Chen:2006sd} and also
other backgrounds~\cite{ListSUSYOthers}. In particular, we
emphasize that Pati-Salam like models, the only models that can
realize all the Yukawa couplings at the stringy tree level, have
been constructed systematically in Type IIA theory on the
$\mathbf{T^6/(\Z_2\times \Z_2)}$ orientifold~\cite{CLL,
Chen:2006sd}. The Standard Model (SM) fermion masses and mixings
can be generated and the gauge coupling unification can be
realized in one of these models~\cite{Chen:2007px, Chen:2007zu},
however, we are not able to stabilize the modulus fields in this
model.

Although some of the complex structure moduli (in Type IIA
picture) and the dilaton field might be stabilized due to the
gaugino condensation in the hidden sector in some models (for
example, see Ref.~\cite{CLW}), the stabilization of all moduli is
still a big challenge.  Important progresses have been made by
introducing background fluxes. In Type IIB theory, the RR fluxes
and NSNS fluxes generate a superpotential~\cite{GVW} that depends
on the dilaton and complex structure moduli, and then stabilize
these moduli dynamically~\cite{Giddings:2001yu, Kachru:2002sk}.
With non-perturbative effects, one can further determine the
K\"ahler moduli~\cite{Kachru:2003aw}. For model building in such
setup, the RR and NSNS fluxes contribute to large positive
D3-brane charges due to the Dirac quantization~\cite{CU, BLT}.
Thus, they modify the global RR tadpole cancellation conditions
significantly and impose strong constraints on the consistent
model building~\cite{MS, CL, Cvetic:2005bn, Kumar:2005hf,
Chen:2005cf}. Including metric fluxes to RR and NSNS fluxes in
Type IIA theory~\cite{Grimm:2004ua, Villadoro:2005cu,
Camara:2005dc}, we can stabilize the moduli in supersymmetric AdS
vacua and relax the RR tadpole cancellation
conditions~\cite{Camara:2005dc, Chen:2006gd}. Interestingly, by
relaxing the RR tadpole cancellation conditions, we can construct
semi-realistic Type IIA~\cite{Chen:2006gd, Chen:2006ip} Pati-Salam
flux models capable of explaining the SM fermion masses and
mixings. However, these models are in the AdS vacua and contain
chiral exotic particles that are difficult to be decoupled.

Including the non-geometric and S-dual
fluxes~\cite{Shelton:2005cf, Aldazabal:2006up, Villadoro:2006ia}
on the Type IIB toroidal orientifolds, the closed string moduli
can be stabilized while the RR tadpole cancellation conditions can
be relaxed elegantly in the supersymmetric Minkowski
vacua~\cite{Chen:2007af}, and the corresponding realistic IIB
Pati-Salam flux models were constructed~\cite{Chen:2007af}.
However, the models in Ref.~\cite{Chen:2007af} contain the
Freed-Witten anomaly \cite{Freed:1999vc} due to its strong
constraints on model building. Interestingly, the Freed-Witten
anomaly can be cancelled by introducing additional
D-branes~\cite{Cascales:2003zp}. In particular, these additional
D-branes do not change the major properties of the D-brane models
such as the four-dimensional $N=1$ supersymmetry and the chiral
particle spectra~\cite{Cascales:2003zp}. Then we demonstrated a
realistic Pati-Salam flux model in a supersymmetric Minkowski
vacuum with the RR tadpole cancellation conditions relaxed and the
Freed-Witten anomaly free conditions satisfied
elegantly~\cite{Chen:2008ht}. Unfortunately, a mistaken flux
algebra in Ref.~\cite{Aldazabal:2006up} made all of the
interesting flux models~\cite{Chen:2008ht} gone with the wind.

Concisely, if we want to stabilize the string moduli, one of the
most important questions in the realistic intersecting D-brane
model building is: {\it Are there supersymmetric Minkowski vacua
with flux compatifications where the moduli can be stabilized and
the RR tadpole cancellation conditions can be relaxed elegantly?}
During the last few years, we have been struggling with the search
for supersymmetric Minkowski vacua with the RR, NSNS, metric,
non-geometric and S-dual fluxes~\cite{Shelton:2005cf,
Aldazabal:2006up, Villadoro:2006ia, Aldazabal:2008zza}
compactified on Type IIB toroidal orientifolds. However, we did
not find any interesting flux vacuum. Therefore we conjecture that
{\it in the generic supsersymmetric Minkowski vacua with flux
compatifications, at least one of the flux contributions to the
seven-brane and D3-brane tadpoles will be positive and then their
tadpole cancellation conditions can not be relaxed if the moduli
are stabilized properly.} In other words, we need to construct
realistic flux models in AdS vacua and then lift them to Minkowski
vacua similar to the KKLT mechanism~\cite{Kachru:2003aw}. A
parallel discussion of no-go theorems for dS vacua on supergravity
algebras with generic fluxes can be found in
\cite{deCarlos:2009fq}.

In this paper, we first review the flux algebra and flux
constraint equations as well as the intersecting D-brane model
building setup on the Type IIB toroidal $\mathbf{T^6}$
orientifolds with the RR, NSNS, metric, non-geometric and S-dual
flux compactifications~\cite{Shelton:2005cf, Aldazabal:2006up,
Villadoro:2006ia,  Aldazabal:2008zza}. We simplify the fluxes
reasonably, discuss the corresponding superpotential, consider the
necessary conditions for supsersymmetric Minkowski vacua and
present all the concrete flux constaint equations. Because
$\mathbf{T^6}$ is factorized as $\mathbf{T^{6}} = \mathbf{T^2}
\times \mathbf{T^2} \times \mathbf{T^2}$, the real parts of the
dilaton, K\"ahler and complex structure moduli must be positive
real numbers when they are stabilized by the fluxes. We show that
we can not have positive real parts of these moduli and
negative/zero contributions to the seven-brane and D3-brane
tadpoles simultaneously in supsersymmetric Minkowski vacua, which
results in lack of realistic flux models because these tadpole
cancellation conditions are not relaxed. The seven-brane and
D3-brane tadpole cancellation conditions can be relaxed elegantly
in supsersymmetric AdS vacua, and we present a concrete Pati-Salam
model as well as its particle speactrum. The lifting from the AdS
vacua to the Minkowski/dS vacua is a great challenge and
definitely needs further studies.

This paper is organized as what follows. In Section II, we briefly
review the flux algebra and flux constraint equations. In Section
III, we briefly review the intersecting D-brane model building. In
Section IV, we reasonably simplify the fluxes, discuss the
necessary conditions for supsersymmetric Minkowski vacua, and
present all the flux constraint equations in details. We show that
there are no consistent supsersymmetric Minkowski vacua with flux
compatifications where all the seven-brane and D3-brane tadpole
cancellation conditions can be relaxed In Section V. We discuss
the supsersymmetric AdS vacua with flux compactifications and
present a semi-realistic Pati-Salam model in Section VI. The
Conclusion is given in Section VII.





\section{Flux Algebra and Flux Constraint Equations}

Let us consider Type IIB orientifold compactifications on
$\mathbf{T}^6/[\Omega(-1)^{F_L}\sigma]$, where $\Omega$ is the
worldsheet parity, $(-1)^{F_L}$ is the left-mover spacetime
fermionic number, and $\sigma$ is the involution. From the metric
of the internal torus, we can introduce the complex structure
moduli $U_k$ and K\"ahler moduli $T_k$, $k=1,2,3$. In addition,
the axion-dilaton modulus $S$ is given by $S=e^{-\phi}+i C_0$ with
$C_0$ the RR 0-form. The K\"ahler potential is
\begin{equation}
K=-\log(S+\bar S)-\sum_{k=1}^3 [\log(U_k+\bar U_k) + \log(T_k+\bar
T_k)].
\end{equation}

We can then introduce the non-trivial RR and NSNS 3-form fluxes,
$F_3$ and $H_3$. These fluxes deform the moduli space and give a
superpotential in the four-dimensional space as \cite{GVW}
\begin{equation}
\mathcal{W}=\int (F_3-iSH_3)\wedge\omega_3.
\end{equation}
The fluxes induce a D3-brane charge through the Chern-Simons
coupling as \cite{Aldazabal:2008zza}
\begin{equation}
\frac{1}{2\cdot 3!}\tilde{F}^{mnp}H_{mnp}=N_{D3/O3},
\label{D3tad-1}
\end{equation}
where $\tilde{F}^{mnp}\equiv
\frac{1}{3!}\epsilon^{ijkopq}F_{opq}$, $i,j,k,m,n,o,p,q$ go from 1
to 6.

The 3-form fluxes are not enough to match the superpotentials of
Type IIA and Type IIB compactifications under the T-duality,
therefore we need to introduce additional fluxes. Consider the IIB
NSNS 3-form flux $H_3$ acted under the T-duality
\cite{Shelton:2005cf, Aldazabal:2006up}:
\begin{equation}
H_{mnp} \stackrel{{\rm T}_m}{\longleftrightarrow} -\omega^m_{np}
\stackrel{{\rm T}_n}{\longleftrightarrow} -Q_p^{mn} \stackrel{{\rm
T}_p}{\longleftrightarrow} R^{mnp},
\end{equation}
where $\omega$, $Q$, and $R$ are introduced as geometric and
non-geometric fluxes. Furthermore, to recover the $SL(2,\Z)$
S-duality in Type IIB theory we introduce one more non-geometric
flux $P$ \cite{Aldazabal:2006up, Aldazabal:2008zza}.

Starting from a general magnetized D9-brane in Type I
compactification on $\mathbf{T}^6$ with ``wrapping numbers''
$(n_1, m_1)\times (n_2,m_2)\times (n_3,m_3)$, under T-daulity this
is corresponding to the D7$_k$-branes in Type IIB. For simplicity,
we also assume an underline $\Z_2\times \Z_2$ symmetry.
Considering the case of extended $(p,q)$ 7-branes, the flux
consistent conditions can be summarized as what
follows~\cite{Aldazabal:2008zza}:
\begin{itemize}

\item Antisymmetry of Commutators
\begin{align}
&{Q}^{ab}_p {P}^{pc}_m-{P}^{ab}_p {Q}^{pc}_m=0\ ,\label{antis1} \\
&-{Q}^{ab}_p {\tilde H}^{clp}+{P}^{ab}_p\,{\tilde F}^{clp}+
{\tilde
H}^{pab}\, {Q}^{cl}_p -{\tilde F}^{pab}{P}^{cl}_p= 0 \ ,\label{antis2} \\
&(\mathcal{F}^I_k)^{ap}(q^I_k Q^{cd}_p-p^I_k P^{cd}_p)= 0\
.\label{antis3}
\end{align}

\item Jocobi Identities

\begin{itemize}

\item Bianchi Identities
\begin{align}
&{Q}^{[ab}_p{Q}^{c]p}_l = 0\ , \label{bian1}\\
&{P}^{[ab}_p{P}^{c]p}_l  = 0\ ,\label{bian2} \\
&{Q}^{[ab}_p{P}^{c]p}_l={P}^{[ab}_p{Q}^{c]p}_l  =
0\,\label{bian3}\\
& Q^{l[a}_{p}\tilde{H}^{bc]p} + Q^{[ab}_{p}\tilde{H}^{c]lp} -
P^{l[a}_{p}\tilde{F}^{bc]p} - P^{[ab}_{p}\tilde{F}^{c]lp} =0.
\label{bian4}
\end{align}

\item Seven-Brane Tadpoles
\begin{align}
& (QF_3)_k=-\sum_{I}
(p_k^I)^2 d_k^I, \label{tad1}\\
& (PH_3)_k=-\sum_{I}
(q_k^I)^2 d_k^I, \label{tad2}\\
& (PF_3+QH_3)_k =-2\sum_{I} p_k^Iq_k^I d_k^I ,\label{tad3}
\end{align}
where $(AB_3)_k = [AB]_{pm}=\frac{1}{2}A^{ab}_{[p} B_{m]ab}$,
$m=p-3=k$.

\item Freed-Witten Anomalies
\begin{align}
Q^{[ab}_p(c^k_I)^{c]p}=0\ ,\label{fw1} \\
P^{[ab}_p(c^k_I)^{c]p}=0\ , \label{fw2}
\end{align}
\end{itemize}
where $c_I^k$ and $d_k^I$ are coefficients of wrapping numbers
of the $I$-th
stack of seven-branes, and will be discussed in details below.

\end{itemize}

The individual items of the fluxes are marked into conventional
notations \cite{Aldazabal:2006up, Aldazabal:2008zza}. In brief, in
the Type IIB picture the fluxes contain elements as $F\supset
\{m,q_k,e_k,e_0 \}$, $H\supset \{\bar h_0,\bar a_k,a_k,h_0 \}$,
$Q\supset \{h_k,b_{ij},\bar b_{ij},\bar h_k \}$, and $P\supset
\{f_k,g_{ij},\bar g_{ij},\bar f_k \}$. The expansions of the flux
constraints by the flux elements can be summarized as
\begin{itemize}

\item Antisymmetry of Commutators

From Eq.~(\ref{antis1}) $QP-PQ=0$, we obtain
\begin{eqnarray}
\bar{b}_{ij}f_{k}+b_{kk}g_{jj}-b_{jj}g_{kk}-\bar{g}_{ij}h_{k} &=&0
\ ,\label{qppq1}\\
b_{ik}\bar{g}_{ij}+b_{jj}\bar{g}_{jk}-\bar{b}_{ij}g_{ik}-\bar{b}_{jk}
g_{jj}&=&0  \ ,\label{qppq2}\\
b_{jk}\bar{g}_{ij}+b_{jj}\bar{g}_{ik}-\bar{b}_{ik}g_{jj}-\bar{b}_{ij}g_{jk}&=&0
\ ,
\label{qppq3}\\
b_{jj}\bar{f}_{k}+\bar{b}_{kk}\bar{g}_{ij}-\bar{b}_{ij}\bar{g}_{kk}-\bar{h}_{k}g_{jj}&=&0
\ ,
\label{qppq4}\\
b_{kj}\bar{g}_{kk}+\bar{h}_kf_{j}-\bar{b}_{kk}g_{kj}-\bar{f}_k
h_j&=&0  \ ,
\label{qppq5}\\
\bar{b}_{ik}f_{j}+b_{kj}g_{jk}-b_{jk}g_{kj}-\bar{g}_{ik}h_{j} &=&0
\ , \label{qppq6}\\
b_{kk}f_j+b_{kj}f_{k}-g_{kk}h_{j}-g_{kj}h_{k}&=&0 \ ,
\label{qppq7}\\
\bar{b}_{jk}f_{j}+b_{kj}g_{ik}-b_{ik}g_{kj}-\bar{g}_{jk}h_{j}&=&0
\ ,
\label{qppq8}\\
b_{kk}\bar{g}_{kj}+\bar h_{j}f_{k}-\bar{b}_{kj}g_{kk}-\bar{f}_{j}
h_{k}&=&0  \ , \label{qppq9}\\
b_{ik}\bar{f}_{j}+\bar{b}_{kj}\bar{g}_{jk}-\bar{b}_{jk}\bar{g}_{kj}-\bar{h}_{j}g_{ik}&=&0
\ ,
\label{qppq10}\\
b_{jk}\bar{f}_{j}+\bar{b}_{kj}\bar{g}_{ik}-\bar{b}_{ik}\bar{g}_{kj}-\bar{h}_{j}g_{jk}
&=&0 \ ,\label{qppq11}\\
\bar{b}_{kk}\bar{f}_{j}+\bar{b}_{kj}\bar{f}_{k}-\bar{g}_{kk}\bar{h}_{j}-\bar{g}_{kj}\bar{h}_{k}&=&0
\ ,\label{qppq12}\\
\bar{b}_{jj}f_{k}+b_{kk}g_{ij}-b_{ij}g_{kk}-\bar{g}_{jj}h_{k} &=&0
\ ,\label{qppq13}\\
b_{ik}\bar{g}_{jj}+b_{ij}\bar{g}_{jk}-\bar{b}_{jk}g_{ij}-\bar{b}_{jj}
g_{ik}&=&0  \ ,\label{qppq14}\\
b_{ij}\bar{g}_{ik}+b_{jk}\bar{g}_{jj}-\bar{b}_{ik}g_{ij}-\bar{b}_{jj}g_{jk}&=&0
\ ,\label{qppq15}\\
b_{ij}\bar{f}_{k}+\bar{b}_{kk}\bar{g}_{jj}-\bar{b}_{jj}\bar{g}_{kk}-\bar{h}_{k}g_{ij}&=&0
\ . \label{qppq16}
\end{eqnarray}


From Eq.~(\ref{antis2}) $-Q\tilde H+P\tilde F+\tilde H Q-\tilde F
P=0$, we obtain
\begin{eqnarray}
a_kb_{ik}+a_ib_{kk}+\bar{g}_{jk} e_0+e_k g_{ik}+e_i
g_{kk}+\bar{b}_{jk}h_0-\bar{a}_jh_k+f_kq_j &=&0 \ ,
\label{qhpf1}\\
(\bar{a}_i b_{ii}+\bar{f}_i e_0+e_i\bar{g}_{ii}+\bar{h}_0h_i)
+\bar{h}_i h_0+ a_i\bar{b}_{ii}+
mf_i-g_{ii}q_i &=&0 \ ,\label{qhpf2}\\
(\bar{a}_i b_{ik}-a_j\bar{b}_{jk}+e_i\bar{g}_{ik}+g_{jk}q_j)
-\bar{a}_jb_{jk}+a_i\bar{b}_{ik}-e_j\bar{g}_{jk}-g_{ik}q_i&=&0\ ,
\label{qhpf3}\\
\bar{a}_k\bar{b}_{jk}+\bar{a}_j\bar{b}_{kk}+b_{ik}\bar{h}_0
-a_i\bar{h}_k-e_i\bar{f}_k+mg_{ik}-\bar{g}_{kk}q_j-\bar{g}_{jk}q_k&=&0
\ . \label{qhpf4}
\end{eqnarray}


From Eq.~(\ref{antis3}) for $k\not =a$, $i=1,2,3$, we obtain
\begin{eqnarray}
\frac{m^I_{a}}{n^I_a}(f_ap^I_k-h_aq^I_k) &=&0  \ ,
\label{fqQ1}\\
\frac{m^I_{a}}{n^I_a}(g_{ia}p^I_k-b_{ia}q^I_k) &=&0  \ ,
\label{fqQ2}\\
\frac{m^I_{a}}{n^I_a}(\bar{g}_{ia}p^I_k-\bar{b}_{ia}q^I_k)&=&0 \ ,
\label{fqQ3}\\
\frac{m^I_{a}}{n^I_a}(\bar{f}_ap^I_k-\bar{h}_aq^I_k) &=&0 \ ,
\label{fqQ4}
\end{eqnarray}


\item Jocobi Identities

\begin{itemize}

\item Bianchi Identities

From Eq.~(\ref{bian1}) $QQ=0$, we obtain
\begin{eqnarray}
-b_{ii}b_{jk}+\bar{b}_{ki}h_k+h_i\bar{b}_{kk}-b_{ji}b_{ik} &=&0 \
, \label{qq1}\\
-\bar{b}_{ii}\bar{b}_{jk}+b_{ki}\bar{h}_k+\bar{h}_ib_{kk}-\bar{b}_{ji}
\bar b_{ik}&=&0  \ ,
\label{qq2}\\
-b_{ii}\bar{b}_{ij}+\bar{b}_{ji}b_{jj}+h_i\bar{h}_j-b_{ki}\bar{b}_{kj}&=&0
\ , \label{qq3}\\
\bar{b}_{ii}b_{ij}-b_{ji}\bar{b}_{jj}+h_i\bar{h}_j-b_{ki}\bar{b}_{kj}&=&0
\  . \label{qq4}
\end{eqnarray}

From Eq.~(\ref{bian2}) $PP=0$, we obtain
\begin{eqnarray}
-g_{ii}g_{jk}+\bar{g}_{ki}f_k+f_i\bar{g}_{kk}-g_{ji}g_{ik}  &=&0 \
, \label{pp1}\\
-\bar{g}_{ii}\bar{g}_{jk}+g_{ki}\bar{f}_k+\bar{f}_ig_{kk}-\bar{g}_{ji}
\bar g_{ik}  &=&0   \ ,
\label{pp2}\\
-g_{ii}\bar{g}_{ij}+\bar{g}_{ji}g_{jj}+f_i\bar{f}_j-g_{ki}\bar{g}_{kj}
&=&0   \ ,
\label{pp3}\\
\bar{g}_{ii}g_{ij}-g_{ji}\bar{g}_{jj}+f_i\bar{f}_j-g_{ki}\bar{g}_{kj}
&=&0   \ . \label{pp4}
\end{eqnarray}

From Eq.~(\ref{bian3}) $QP=0$, we obtain
\begin{eqnarray}
b_{kk}\bar g_{kj} -h_k \bar f_j - \bar b_{jk}g_{jj} + b_{ik}\bar
g_{ij} &=&0 \ , \label{qp1}\\
b_{kk} g_{ij} -h_k \bar g_{jj} - \bar b_{jk} f_j + b_{ik} g_{kj}
&=&0
\ , \label{qp2}\\
\bar b_{kk} \bar g_{ij} - \bar h_k g_{jj} - b_{jk} \bar f_j + \bar
b_{ik} \bar g_{kj}   &=&0   \ ,
\label{qp3}\\
\bar b_{kk} g_{kj} - \bar h_k f_j - b_{jk} \bar g_{jj} + \bar
b_{ik} g_{ij} &=&0 \ . \label{qp4}
\end{eqnarray}

From Eq.~(\ref{bian3}) $PQ=0$, we obtain
\begin{eqnarray}
g_{kk}\bar b_{kj} - f_k \bar h_j - \bar g_{jk}b_{jj} + g_{ik}\bar
b_{ij} &=&0 \ ,
\label{pq1}\\
g_{kk} b_{ij} - f_k \bar b_{jj}  - \bar g_{jk} h_j + g_{ik} b_{kj}
&=&0 \ ,
\label{pq2}\\
\bar g_{kk} \bar b_{ij} - \bar f_k b_{jj}  - g_{jk} \bar h_j +
\bar g_{ik} \bar b_{kj}  &=&0   \ ,
\label{pq3}\\
\bar g_{kk} b_{kj} - \bar f_k h_j  - g_{jk} \bar b_{jj} + \bar
g_{ik} b_{ij} &=&0   \ . \label{pq4}
\end{eqnarray}

From Eq.~(\ref{bian4}) $Q^[\tilde{H}^]+ Q^[\tilde{H}^]-
P^[\tilde{F}^]-P^[\tilde{F}^]=0$, we obtain
\begin{eqnarray}
a_kb_{ik}+a_ib_{kk}+\bar{g}_{jk}
e_0+e_k g_{ik}+e_i g_{kk}+\bar{b}_{jk}h_0-\bar{a}_jh_k+f_kq_j &=&0 \label{qhqhpfpf1},\\
\bar{a}_k\bar{b}_{jk}+\bar{a}_j\bar{b}_{kk}+b_{ik}\bar{h}_0
-a_i\bar{h}_k-e_i\bar{f}_k+mg_{ik}-\bar{g}_{kk}q_j-\bar{g}_{jk}q_k&=&0
\label{qhqhpfpf2}, \\
(\bar a_i b_{ii} + \bar h_0 h_i + e_0 \bar f_i + e_i \bar g_{ii})
-a_j \bar b_{ji} + \bar a_k b_{ki} + e_k \bar g_{ki} + q_j g_{ji}
&&\nonumber \\
+(h_0 \bar h_i + a_i \bar b_{ii} + m f_i -q_i g_{ii}) + a_k \bar
b_{ki} -\bar a_j b_{ji} -e_j \bar g_{ji} - q_k g_{ki} &=& 0.
\label{qhqhpfpf4}
\end{eqnarray}

\item Seven-Brane Tadpole Constraint Equations

Let $(n^I_i, m^I_i)$, $i=1,2,3$ the wrapping numbers, then
$d_i^I=-n_i^I m_j^I m_k^I$, $i\neq j\neq k$. From
Eq.~(\ref{tad1}), we obtain
\begin{eqnarray}
-\sum_I (p^I_i)^2d_i^I +\frac{1}{2}[mh_i -e_0 \bar{h}_i -\sum_j
(q_j b_{ji}+e_j\bar{b}_{ji})]=0 \ \label{D7tad1}~.~\
\end{eqnarray}

From Eq.~(\ref{tad2}), we obtain
\begin{eqnarray}
-\sum_I (q^I_i)^2d_i^I +\frac{1}{2}[h_0 \bar f_i - \bar h_0 f_i -
\sum_j (\bar a_j g_{ji} - a_j \bar g_{ji} )] =0 \label{D7tad2}~.~\
\end{eqnarray}

From Eq.~(\ref{tad3}), we obtain
\begin{eqnarray}
-2\sum_I (p^I_iq^I_i)d^I_i &+& \frac{1}{2}[\bar h_0 h_i-\bar h_i
h_0 +\sum_j (\bar a_j b_{ji} - a_j \bar b_{ji} ) \nonumber
\\ & +&e_0\bar f_i-mf_i +\sum_j (q_j g_{ji} + e_j \bar g_{ji}
)]=0~.\label{D7tad3}
\end{eqnarray}

\item D3-Brane RR Tapole Constraint Equation

From Eq.~(\ref{D3tad-1}), we obtain
\begin{equation}
\sum_I n_1^In_2^In_3^I + \frac{1}{2}[m h_0 - e_0 \bar h_0 + \sum_i
(q_i a_i + e_i \bar a_i) ] = 16 ~.~ \label{D3tad}
\end{equation}

\item Freed-Witten Anomalies

From Eq.~(\ref{fw1}), for $i\not =j\not =k$, $a=1,2,3$, we obtain
\begin{eqnarray}
(h_jm^I_{j}n^I_i+h_im^I_{i}n^I_j)n_k &=&0 \ ,
\label{fpQ1}\\
(b_{aj}m^I_{j}n^I_i+b_{ai}m^I_{i}n^I_j)n_k &=&0  \ ,
\label{fpQ2}\\
(\bar{b}_{aj}m^I_{j}n^I_i+\bar{b}_{ai}m^I_{i}n^I_j)n_k &=&0 \
, \label{fpQ3}\\
(\bar{h}_jm^I_{j}n^I_i+\bar{h}_im^I_{i}n^I_j)n_k &=&0 \ .
\label{fpQ4}
\end{eqnarray}

From Eq.~(\ref{fw2}), for $i\not =j\not =k$, $a=1,2,3$, we obtain
\begin{eqnarray}
(f_jm^I_{j}n^I_i+f_im^I_{i}n^I_j)n_k &=&0 \ , \label{fqP1}\\
(g_{aj}m^I_{j}n^I_i+g_{ai}m^I_{i}n^I_j)n_k &=&0  \ ,
\label{fqP2}\\
(\bar{g}_{aj}m^I_{j}n^I_i+\bar{g}_{ai}m^I_{i}n^I_j)n_k &=&0 \ ,
\label{fqP3}\\
(\bar{f}_jm^I_{j}n^I_i+\bar{f}_im^I_{i}n^I_j)n_k &=&0 \ .
\label{fqP4}
\end{eqnarray}

\end{itemize}
\end{itemize}

\section{Intersecting D-Brane Model Building on Type IIB Orientifold}

We now consider the Type IIB string theory compactified on a
$\mathbf{T^6}$ orientifold where $\mathbf{T^{6}}$ is a six-torus
factorized as $\mathbf{T^{6}} = \mathbf{T^2} \times \mathbf{T^2}
\times \mathbf{T^2}$ whose complex coordinates are $z_i$, $i=1,\;
2,\; 3$ for the $i$-th two-torus, respectively~\cite{CU, BLT,
Cvetic:2005bn}. The orientifold projection is implemented by
gauging the symmetry $\Omega R$, where $\Omega$ is world-sheet
parity, and $R$ is given by
\begin{eqnarray}
R: (z_1,z_2,z_3) \to (-z_1, -z_2, -z_3)~.~\,    \label{orientifold}
\end{eqnarray}
Thus, the model contains 64 O3-planes.
In order to cancel the negative RR charges from these
O3-planes, we introduce the magnetized
D(3+2n)-branes which are filling up the
four-dimensional Minkowski space-time and wrapping
2n-cycles on the compact manifold. Concretely, for one stack
of $N_a$ D-branes wrapped $m_a^i$ times on the  $i$-th
two-torus $\mathbf{T^2_i}$, we turn on $n_a^i$ units of magnetic fluxes
$F^i_a$ for the center of mass $U(1)_a$ gauge factor on $\mathbf{T^2_i}$,
such that
\begin{eqnarray}
m_a^i \, \frac 1{2\pi}\, \int_{T^2_{\,i}} F_a^i \, = \, n_a^i ~,~\,
\label{monopole}
\end{eqnarray}
where $m_a^i$ can be half integer for tilted two-torus.
Then, the D9-, D7-, D5- and D3-branes contain 0, 1, 2 and 3 vanishing
$m_a^i$s, respectively. Introducing for the $i$-th two-torus
the even homology classes $[{\bf 0}_i]$ and $[{\bf T}^2_i]$ for
the point and two-torus, respectively, the vectors of the RR
charges of the $a$ stack of D-branes and its image are
\begin{eqnarray}
 && [{ \Pi}_a]\, =\, \prod_{i=1}^3\, ( n_a^i [{\bf 0}_i] + m_a^i [{\bf T}^2_i] ),
 \nonumber\\&&
[{\Pi}_a']\, =\, \prod_{i=1}^3\, ( n_a^i [{\bf 0}_i]- m_a^i [{\bf T}^2_i] )~,~
\label{homology class for D-branes}
\end{eqnarray}
respectively.
The ``intersection numbers'' in Type IIA language, which determine
the chiral massless spectrum, are
\begin{eqnarray}
I_{ab}&=&[\Pi_a]\cdot[\Pi_b]=\prod_{i=1}^3(n_a^im_b^i-n_b^im_a^i)~.~
\label{intersections}
\end{eqnarray}
Moreover, for a stack of $N$ D(2n+3)-branes whose homology classes
on $\mathbf{T^{6}}$ is (not) invariant under $\Omega R$, we obtain
a ($U(N)$) $USp(2N)$  gauge symmetry with three (adjoint)
anti-symmetric chiral superfields due to the orbifold projection.
The physical spectrum is presented in Table \ref{spectrum}.

\begin{table}[t]
\caption{General spectrum  for magnetized D-branes on the  Type IIB
${\mathbf{T^6}}$ orientifold. }
\renewcommand{\arraystretch}{1.25}
\begin{center}
\begin{tabular}{|c|c|}
\hline {\bf Sector} & {\bf Representation}
 \\
\hline\hline
$aa$   & $U(N_a)$ vector multiplet  \\
       & 3 adjoint multiplets  \\
\hline
$ab+ba$   & $I_{ab}$ $(N_a,{\overline{N}}_b)$ multiplets  \\
\hline
$ab'+b'a$ & $I_{ab'}$ $(N_a,N_b)$ multiplets \\
\hline $aa'+a'a$ &$\frac 12 (I_{aa'} -  I_{aO3})\;\;$
symmetric multiplets \\
          & $\frac 12 (I_{aa'} +  I_{aO3}) \;\;$
anti-symmetric multiplets \\
\hline
\end{tabular}
\end{center}
\label{spectrum}
\end{table}

The flux models on Type IIB orientifolds with four-dimensional
$N=1$ supersymmetry  are primarily constrained by the RR tadpole
cancellation conditions that will be given later, the
four-dimensional $N=1$ supersymmetry condition, and the K-theory
anomaly free conditions. For the D-branes with world-volume
magnetic field $F_a^i={n_a^i}/({m_a^i\chi_i})$ where $\chi_i$ is
the area of $\mathbf{T^2_i}$ in string units,  the condition to
preserve the four-dimensional $N=1$ supersymmetry
is~\cite{Cvetic:2005bn}
\begin{eqnarray}
\sum_i \left(\tan^{-1} (F_a^i)^{-1} +
{\theta (n_a^i)} \pi \right)=0 ~~~{\rm mod}~ 2\pi~,~\,
\end{eqnarray}
where ${\theta (n_a^i)}=1$ for $n_a^i < 0$ and
${\theta (n_a^i)}=0$ for $n_a^i \geq 0$.
The K-theory anomaly free conditions are
\begin{eqnarray}
&&  \sum_a N_a m_a^1 m_a^2 m_a^3 =  \sum_a N_a m_a^1 n_a^2 n_a^3
= \sum_a N_a n_a^1 m_a^2 n_a^3
\nonumber\\&&
= \sum_a N_a n_a^1 n_a^2 m_a^3
=0 ~~~{\rm mod}~ 2~.~\,
\end{eqnarray}
And the holomorphic gauge kinetic function for a generic stack of
D(2n+3)-branes  is given by~\cite{Cremades:2002te, Lust:2004cx}
\begin{eqnarray}
f_a = {1\over {\kappa_a}}\left(  n_a^1\,n_a^2\,n_a^3\,s-
n_a^1\,m_a^2\,m_a^3\,t_1 -n_a^2\,m_a^1\,m_a^3\,t_2
-n_a^3\,m_a^1\,m_a^2\,t_3\right)~,~\, \label{EQ-GKF}
\end{eqnarray}
where $\kappa_a$ is equal to  1 and 2 for $U(n)$ and $USp(2n)$,
respectively.



In general, this kind of D-brane models possesses Freed-Witten
anomalies~\cite{CU, Freed:1999vc}. In the world-volume of a
generic stack of D-branes we have a $U(1)$ gauge field whose
scalar partner parameterizes the D-brane position in the compact
space. Such kind of $U(1)$'s usually obtain St\"uckelberg masses
by swallowing RR scalar fields and then decouple from the
low-energy spectrum. At the same time these scalars participate in
the cancellations of $U(1)$ gauge anomalies through a generalized
Green-Schwarz mechanism~\cite{Aldazabal:2000dg}.


\section{Simplifying the Flux Constraint Equations and Conditions for
Supersymmetric Minkowski Vacua}

The system with general fluxes is very complicated, therefore, we
intend to simply the system by some isotropy conditions. First we
assume the three complex structure moduli the same. For
simplicity, we redefine the dilaton $S$, the three K\"ahler moduli
$T_i$, and the three complex structure moduli $U_i$ as
\begin{equation}
S \equiv -i\sigma,~~~ T_i \equiv -i\tau_i,~~~ U_i \equiv U \equiv -i\rho~.~
\end{equation}
Because the real parts of $S$, $T_i$, and $U_i$ must be positive
real numbers, the imaginary parts of $\sigma$, $\tau_i$ and $\rho$
must be positive as well. The fluxes can be simplified as
\begin{eqnarray}
&& e_i= e,~~~ q_i=q,~~~  a_i=a,~~~ \bar{a}_i= \bar{a}, \nonumber \\
&& b_{ij}=b_j,~~~ b_{ii}=\beta_i,~~~ \bar b_{ij}=\bar b_j,~~~
\bar b_{ii}=\bar\beta_i\\
&& g_{ij}=g_j,~~~ g_{ii}=\gamma_i,~~~ \bar g_{ij}=\bar g_j,~~~
\bar g_{ii}=\bar\gamma_i.
\end{eqnarray}
The corresponding superpotential including the fluxes is
\begin{eqnarray}
\mathcal{W} &=& e_0 +3e\rho +3q\rho^2-m \rho^3 \nonumber \\
&+& \sigma[h_0 +3a\rho -3\bar a \rho^2-\bar h_0 \rho^3 ] \\
&+& \sum_i\tau_i[-h_i +(2b_i+\beta_i)\rho -(2\bar b_i+\beta_i)
\rho^2 + \bar h_i\rho^3] \nonumber  \\
&+& \sigma\sum_i\tau_i[f_i -(2g_i+\gamma_i)\rho +(2\bar
g_i+\gamma_i) \rho^2 - \bar f_i\rho^3]. \nonumber
\end{eqnarray}

The K\"ahler moduli are not simplified yet because we want to keep
some degrees of freedoms for model building. For convenience we
assume $\tau \mathbb{F}=\tau_i \mathbb{F}_i$, $\tau_i=k_i\tau$ and
$\mathbb{F}_i=\mathbb{F}/k_i$, where $k_i$ are real and positive
constants depending on the models, and $\mathbb{F}\in \{b,\bar b,
\beta, \bar\beta, f,\bar f, g, \bar g, \gamma, \bar\gamma, h, \bar
h\}$. Thus, the real part of $\tau$ must be also a positive real
number,  and the superpotential $\mathcal{W}$ turns out
\begin{eqnarray}
\mathcal{W} &=& E_1 + \sigma E_2 + \tau E_3 +\sigma\tau E_4\nonumber  \\
&=& e_0 +3e\rho +3q\rho^2-m \rho^3 \nonumber \\
&+& \sigma[h_0 +3a\rho -3\bar a \rho^2-\bar h_0 \rho^3 ] \\
&+& 3\tau[-h +(2b+\beta)\rho -(2\bar b+\beta)
\rho^2 + \bar h\rho^3] \nonumber  \\
&+& 3\sigma\tau[f-(2g+\gamma)\rho +(2\bar g+\gamma) \rho^2 - \bar
f\rho^3]. \nonumber
\end{eqnarray}

If we pursue AdS vacua, then it is required that
\begin{equation}
\frac{\partial \mathcal{W}}{\partial \sigma} =\frac{\partial
\mathcal{W}}{\partial \rho} = \frac{\partial \mathcal{W}}{\partial
\tau}=0.
\end{equation}
If $E_4\neq 0$, it implies
\begin{equation}
\sigma = -\frac{E_3}{E_4},~~~ \tau=-\frac{E_2}{E_4},
\end{equation}

On the other hand, if we look for Minkowski vacua, the additional
condition is $\mathcal{W}=0$, so we can define a new polynomial
$E$ as
\begin{equation}
E=E_1E_4-E_2E_3=0,~~~ \frac{\partial E}{\partial \rho}=0.
\label{Minkowski}
\end{equation}
The function must have a double root $\rho_0$, and its complex
conjugate. Therefore, the function $E=0$ can be written as
\begin{equation}
E=(\rho-\rho_0)^2(\rho-\rho_0^{\ast})^2(A\rho^2+B\rho+C).
\end{equation}

The simplified flux constraint equations are:

\subsubsection{Antisymmetry of Commutators}

From Eq.~(\ref{antis1}) $QP-PQ=0$, we obtain
\begin{eqnarray}
(\ref{qppq1}),(\ref{qppq6}),(\ref{qppq8})&\rightarrow & \bar b f =\bar g h ~,~\label{qppqn1}\\
(\ref{qppq2})=(\ref{qppq3}) &\rightarrow & b\bar g - \bar b g +\beta \bar g -\bar b \gamma =0~,~\label{qppqn2}\\
(\ref{qppq10}),(\ref{qppq11}),(\ref{qppq16})&\rightarrow & b \bar f = g \bar h ~,~\label{qppqn3}\\
(\ref{qppq14})=(\ref{qppq15}) &\rightarrow & b\bar g -\bar b g -\bar\beta g +b \bar\gamma =0 ~,~\label{qppqn4}\\
(\ref{qppq4}) &\rightarrow & \beta \bar f -\bar h \gamma + \bar\beta\bar g -\bar b\bar\gamma=0~,~ \label{qppqn5} \\
(\ref{qppq5}) &\rightarrow & f\bar h -\bar f h -\bar\beta g +b \bar\gamma =0~,~ \label{qppqn6}\\
(\ref{qppq7}) &\rightarrow & \beta f - h \gamma + b f -g h=0~,~ \label{qppqn7} \\
(\ref{qppq9}) &\rightarrow & f\bar h -\bar f h +\beta \bar g - \bar b\gamma =0~,~ \label{qppqn8}\\
(\ref{qppq12}) &\rightarrow & \bar\beta\bar f -\bar h\bar\gamma +\bar b\bar f -\bar g\bar h =0~,~ \label{qppqn9}\\
(\ref{qppq13}) &\rightarrow & \bar\beta f -h \bar \gamma +\beta g
-b\gamma =0 ~.~\label{qppqn10}
\end{eqnarray}

From Eq.~(\ref{antis2}) $-Q\tilde H+P\tilde F+\tilde H Q-\tilde F
P=0$, we obtain
\begin{eqnarray}
(\ref{qhpf1}) &\rightarrow& ab+a\beta+e_0\bar g + eg +e\gamma
+\bar b h_0-\bar a h+fq=0 ~,~\label{qhfpN1}\\
(\ref{qhpf2}) &\rightarrow& \bar a\beta+a\bar\beta+e_0\bar f
+e\bar\gamma +h_0\bar h +\bar h_0 h + mf -\gamma q=0 ~,~\label{qhfpN2}\\
(\ref{qhpf3}) &\rightarrow& 0=0 \label{qhfpN3}~,~\\
(\ref{qhpf4}) &\rightarrow& \bar a\bar b+\bar a\bar\beta-e\bar f +
b \bar h_0-a\bar h + mg-\bar g q-\bar \gamma q=0 ~.~\label{qhfpN4}
\end{eqnarray}

From Eq.~(\ref{antis3}), $i\neq k$, we obtain
\begin{eqnarray}
&&\frac{m^I_{i}}{n^I_i}(f_ip^I_k - h_i q^I_k) =0 ,  \label{anti3-m1}\\
&&\frac{m^I_{i}}{n^I_i}(g_ip^I_k - b_i q^I_k)=0,
~~~\frac{m^I_{i}}{n^I_i}(\gamma_ip^I_k - \beta_i q^I_k) =0 ,\label{anti3-m2}\\
&&\frac{m^I_{i}}{n^I_i}(\bar g_ip^I_k - \bar b_i q^I_k)=0,
~~~\frac{m^I_{i}}{n^I_i}(\bar\gamma_ip^I_k - \bar\beta_i q^I_k) =0,\label{anti3-m3}\\
&&\frac{m^I_{i}}{n^I_i}(\bar f_ip^I_k - \bar h_i q^I_k)=0
\label{anti3-m4}.
\end{eqnarray}

\subsubsection{Jocobi Identities}

\begin{itemize}

\item Bianchi Identities

From Eq.~(\ref{bian1}) $QQ=0$, we obtain
\begin{eqnarray}
(\ref{qq1}) &\rightarrow & -\beta b+\bar{b}h +h\bar{\beta}-bb =0 ~,~\label{qqN1}\\
(\ref{qq2}) &\rightarrow &-\bar{\beta}\bar{b} +b\bar{h}+\bar{h}\beta-\bar{b}\bar b =0 ~,~\label{qqN2}  \\
(\ref{qq3}),(\ref{qq4})  &\rightarrow & h\bar h = b \bar b~.~
\label{qqN3}
\end{eqnarray}

From Eq.~(\ref{bian2}) $PP=0$, we obtain
\begin{eqnarray}
(\ref{pp1}) &\rightarrow & -\gamma g+\bar{g}f +f\bar{\gamma}-gg =0 ~,~\label{ppN1}\\
(\ref{pp2}) &\rightarrow &-\bar{\gamma}\bar{g} +g\bar{f}+\bar{f}\gamma-\bar{g}\bar g =0 ~,~\label{ppN2}  \\
(\ref{pp3}),(\ref{pp4})  &\rightarrow & f\bar f = g \bar g~.~
\label{ppN3}
\end{eqnarray}

From Eq.~(\ref{bian3}) $QP=0$, we obtain
\begin{eqnarray}
(\ref{qp1}) &\rightarrow & \beta\bar g - h\bar f -\bar b\gamma +
b\bar g=0,
~ \oplus(\ref{qppqn2}) \rightarrow \bar f h= \bar b g ~,~\label{qpN1} \\
(\ref{qp2}) &\rightarrow & -\beta g+ \bar{b}f +h\bar{\gamma}-bg =0 ~,~\label{qpN2}\\
(\ref{qp3}) &\rightarrow &-\bar{\beta}\bar{g} +b\bar{f}+\bar{h}\gamma-\bar{b}\bar g =0 ~,~\label{qpN3} \\
(\ref{qp4}) &\rightarrow & \bar\beta g - \bar h f -b\bar\gamma +
\bar b g=0, ~ \oplus(\ref{qppqn4}) \rightarrow f \bar h  = b \bar
g ~.~\label{qpN4}
\end{eqnarray}

From Eq.~(\ref{bian3}) $PQ=0$, we obtain
\begin{eqnarray}
(\ref{pq1}) &\rightarrow & \bar b\gamma - f\bar h -\beta\bar g +
\bar b g=0, ~ \oplus(\ref{qppqn2}) \rightarrow f\bar h = b\bar g ~,~\label{pqN1} \\
(\ref{pq2}) &\rightarrow &  b\gamma - f\bar\beta -h\bar g + b g=0 ~,~\label{pqN2}\\
(\ref{pq3}) &\rightarrow & \bar b\bar\gamma - \bar f \beta -\bar h g + \bar b \bar g=0 ~,~\label{pqN3} \\
(\ref{pq4}) &\rightarrow &  b\bar\gamma - \bar f h -\bar\beta g +
b\bar g=0, ~ \oplus(\ref{qppqn4}) \rightarrow  \bar f h  =  \bar b
g ~.~\label{pqN4}
\end{eqnarray}

From Eq.~(\ref{bian4}) $Q^[\tilde{H}^]+ Q^[\tilde{H}^]-
P^[\tilde{F}^]-P^[\tilde{F}^]=0$, we obtain
\begin{eqnarray}
(\ref{qhqhpfpf1})\rightarrow(\ref{qhfpN1}),~~~
(\ref{qhqhpfpf2})\rightarrow(\ref{qhfpN4}),~~~
(\ref{qhqhpfpf4})\rightarrow(\ref{qhfpN2}).   \nonumber
\end{eqnarray}

\end{itemize}

\section{ Supersymmetric Minkowski Flux Vacua }

\subsection{Scenarios with Flux $P=0$}

It is hard to satisfy the third anti-symmetric conditions in
Eqs.~(\ref{anti3-m1})-(\ref{anti3-m4}) for a model consisting of more
than two stacks of D-branes. Therefore, a reasonable choice is
setting $P=0$, which implies $E_4=0$. Then the superpotential
turns out to be
\begin{equation}
\mathcal{W}=E_1+\sigma E_2+ \tau E_3.
\end{equation}
To obtain Minkowski vacua, we can conclude the conditions as
\begin{equation}
E_1=E_2=E_3=0,~~~\sigma=-\frac{E_1'}{E_2'},~~~
\tau=-\frac{E_1'}{E_3'}.
\end{equation}
The dilaton modulus $\sigma$ is determined by $E_2$ while the
K\"ahler modulus $\tau$ is controlled by $E_3$. For simplicity, we
will temporarily ignore $E_2$ by setting the corresponding fluxes
$H:\{ a, \bar a, h_0, \bar h_0 \}$ zero because they are related
to fewer constraints and can be included independently later. Then
we obtain the following constraints of the fluxes in $E_3$ from
the Bianchi identities
\begin{equation}
b(b+\beta)=h(\bar b+\bar\beta),~~~ \bar h(b+\beta)= \bar b(\bar
b+\bar\beta),~~~ b\bar b =h \bar h.
\end{equation}
Rewritting these flux conditions by introducing two parameters $\xi$
and $\chi$, we obtain
\begin{eqnarray}
h =b{\xi},~~~ b+\beta=h\chi=b\xi\chi,~~~ \bar b =\bar h\xi,~~~
\bar b+\bar\beta=b\chi,  \label{xichi}
\end{eqnarray}
Therefore, $E_3$ can be factorized as
\begin{eqnarray}
E_3&=&3\left[ -h+(b+h\chi)\rho -(\xi\bar h
+b\chi)\rho^2 +\bar h \rho^3\right]\\
&=& 3(\rho-\xi)(\bar h \rho^2-b\chi\rho +b).  \label{E3-fac}
\end{eqnarray}
The complex roots of $E_3$ are
\begin{equation}
\rho_0=\frac{b\chi\pm\sqrt{b^2\chi^2-4b\bar h}}{2\bar h}
~\Rightarrow~ b^2\chi^2<4b\bar h,~~ b\bar h>0,
\end{equation}
and
\begin{equation}
{\rm Re}(\rho_0)=\frac{b\chi}{2\bar h}\equiv R,~~~ {\rm
Im}(\rho_0)= \pm\frac{\sqrt{|b^2\chi^2-4b\bar h|}}{2\bar h}
\equiv\pm I.
\end{equation}
$E_1$ has the same complex roots, so we assume
\begin{eqnarray}
E_1&=& e_0+3e\rho+3q\rho^2-m\rho^3 = (A\rho+B)(\bar
h\rho^2-b\chi\rho +b).
\end{eqnarray}
Comparing the coefficients, we obtain
\begin{eqnarray}
A=-\frac{m}{\bar h}, ~~~ B=\frac{e_0}{b}, ~~~ 3q=\frac{e_0\bar h
}{b}+\frac{mb\chi}{\bar h},~~~ 3e=-\frac{mb}{\bar h }-e_0\chi.
\end{eqnarray}
Then, we have
\begin{eqnarray}
\tau_0&=&-\frac{E_1'}{E_3'}|_{\rho_0}=-\frac{(A\rho_0+B)}{3(\rho_0-\xi)}~,~\\
{\rm Im}(\tau_0)&=&\frac{\pm I}{3[(R-\xi)^2+I^2]}\left(
\frac{e_0}{b}-\frac{mh}{\bar h b} \right) =\frac{\rm
Im(\rho_0)}{3[(R-\xi)^2+I^2]} \frac{( e_0\bar h-mh )}{b\bar h}~.~
\end{eqnarray}

\subsubsection{$(p,q)=(1,0)$}

First we consider the case with only D-branes where $p^I_i=1$ and
$q^I_i=0$. Recall that $b\bar h>0$. The D7-brane tadpole
contribution from the fluxes is
\begin{eqnarray}
N_{flux\rm D7}&=&\frac{1}{2}[mh-e_0\bar h-(2b+\beta)q-(2\bar
b+\bar\beta)e] \nonumber \\
&=&\frac{1}{6}(mh-e_0\bar h)\left( 4-\frac{b\chi^2}{\bar h}
\right).
\end{eqnarray}
In this paper, we denote and emphasize that $N_{flux\rm D7}$ is
the number of D7-branes that we need to introduce for D7-brane
tadpole cancellations due to the flux contributions. In other
words, $-N_{flux\rm D7}$ is the flux contribution to the D7-brane
tadpoles.

We have learned $b^2\chi^2<4b\bar h$ for $\rho_0$ being complex,
so $4-\frac{b\chi^2}{\bar h}  >0$. It turns out that $N_{flux\rm
D7}$ has always negative sign to Im($\tau_0$). Thus, we do not
have supersymmetric Minkowski flux vacua that the D7-brane tadpole
cancellation condition is relaxed and the moduli are stabilized
properly.

\subsubsection{$(p,q)=(0,0)$}

The D7-brane tadpole contribution from the fluxes has to be zero,
therefore
\begin{eqnarray}
N_{flux\rm D7} = \frac{1}{6}(mh-e_0\bar h)\left(
4-\frac{b\chi^2}{\bar h} \right)=0.
\end{eqnarray}
For $\rho_0$ to be a complex number, we must require
 $4-\frac{b\chi^2}{\bar h}>0$.
So we have $mh-e_0\bar h=0$, which implies that $\tau_0$ has only
real part.  Thus, the moduli $\tau_i$ can not be stabilized
properly.

\subsubsection{$p\cdot q\neq 0$}

Finally we consider the general case with $p\cdot q\neq 0$.
Similarly to the $(p,q)=(1,0)$ case, $N_{flux\rm D7}$ has always
negative sign to Im($\tau_0$).  And since $P=0$, the flux
contribution to the NS7-brane tadpole is zero. Thus, we do not
have supersymmetric Minkowski flux vacua with the D7-brane tadpole
cancellation condition relaxed and the moduli stabilized properly.

\subsection{General Discussions for $P\neq0$ }

The constraint condition of the flux $P=0$ is too stringent,
thus, we
consider the cases with $P\neq 0$.  With the fluxes in $E_4$
non-zero, the Bianchi identities in Eqs.~(\ref{ppN1})-(\ref{ppN3}) give
\begin{equation}
g(g+\gamma)=f(\bar g+\bar\gamma),~~~ \bar f(g+\gamma)= \bar g(\bar
g+\bar\gamma),~~~ g\bar g =f \bar f.
\end{equation}
Similarly, we can rewrite these flux conditions by introducing
parameters $\xi'$ and $\chi'$ as
\begin{eqnarray}
f =g{\xi'},~~~ g+\gamma=f\chi'=g\xi'\chi',~~~ \bar g =\bar
f\xi',~~~ \bar g+\bar\gamma=g\chi' ~.~\,   \label{xi'chi'}
\end{eqnarray}
In addition, there are also the Bianchi conditions between fluxes
$P$ and $Q$:
\[  f\bar h=b\bar g,~~~ \bar f h=\bar b g. \]
After factorization and using Eq.~(\ref{qppqn3}),
we can rewrite $E_4$ as follows
\begin{eqnarray}
E_4 =-3(\rho-\xi')(\bar f \rho^2-g\chi'\rho +g)
=-3\frac{g}{b}(\rho-\xi')(\bar h \rho^2-b\chi'\rho +b).
\end{eqnarray}
From the previous subsection we learned that $E_3$ can be
rewritten as $ E_3= 3(\rho-\xi)(\bar h \rho^2-b\chi\rho +b)$, and
$\chi^2-\frac{4\bar h}{b}<0$ for $\rho$ being complex. Pluging
Eqs.~(\ref{xichi}) and (\ref{xi'chi'}) into the antisymmetry
conditions, we find that it must be $\chi'=\chi$. In other words,
$E_3$ and $E_4$ have the same factor. Recall the Minkowski
condition in Eq.~(\ref{Minkowski}), we obtain
\begin{eqnarray}
E&=&E_1E_4-E_2E_3 \nonumber \\
&=& -\frac{3}{b}(\bar h \rho^2-b\chi\rho
+b)[g(\rho-\xi')(e_0+3e\rho+3q\rho^2-m\rho^3) \nonumber
\\ && +b(\rho -\xi) (h_0+3a\rho-3\bar a \rho^2-\bar
h_0\rho^3) ],
\end{eqnarray}
and
\begin{equation}
\sigma =-\frac{E_3}{E_4} =\frac{b(\rho-\xi)}{g(\rho-\xi')},
~~~~\tau=-\frac{E_2}{E_4}.
\end{equation}
Since the coefficients of $E_2$ are real and $\tau_0$ is limited
with $\rho_0$, $E_2$ must also contain the factor $(\bar h
\rho^2-b\chi\rho +b)$. In addition, $E$ must have double roots for
$E'=0$, therefore $E_1$ has the same factor $(\bar h
\rho^2-b\chi\rho +b)$ as well. In summary, $E_1$ and $E_2$ can be
rewritten as
\begin{eqnarray}
E_1= -(\frac{m}{\bar h}\rho-\frac{e_0}{b})(\bar h \rho^2-b\chi\rho
+b),~~~ E_2 =-(\frac{\bar h_0}{\bar h}\rho-\frac{h_0}{b}) (\bar h
\rho^2-b\chi\rho +b),
\end{eqnarray}
with the flux relations
\begin{eqnarray}
&&  q=\frac{1}{3} \left( \frac{e_0 \bar h}{b} +
\frac{mb}{\bar h} \chi \right),
~~~~~ e=-\frac{1}{3}\left( \frac{mb}{\bar h}+e_0\chi \right), \nonumber \\
&&  \bar a= -\frac{1}{3}\left( \frac{h_0\bar h}{b} + \frac{\bar
h_0b}{\bar h} \chi \right), ~~ a=-\frac{1}{3}\left( \frac{\bar h_0
b}{\bar h}+ h_0\chi \right).
\end{eqnarray}

The imaginary parts of moduli $\sigma$ and $\tau$ must have the
same sign as the imaginary part of $\rho$. From the above
equations we obtain
\begin{equation}
\sigma
=\frac{b(\rho-\xi)}{g(\rho-\xi')},~ \Longrightarrow~ 
\frac{b}{g}(\xi-\xi')=\frac{gh-bf}{g^2}>0; \label{sigma-pq}
\end{equation}
\begin{equation}
\tau =\frac{-f\bar h_0\rho +\bar g h_0}{3\bar g(g\rho-f)},
~\Longrightarrow~ -\frac{1}{3g\bar f} (\bar f h_0-f\bar h_0)>0.
\label{tau-pq}
\end{equation}

The antisymmetry constraints from Eqs.~(\ref{qhfpN1}), (\ref{qhfpN2}),
and (\ref{qhfpN4}) can be written as
\begin{eqnarray}
&&(hh_{0} + e_{0}f)(4-\frac{b\chi^2}{\bar h}) = 0, \nonumber  \\
&&(\bar{h}_{0}\bar h+m\bar f) (4-\frac{b\chi^2}{\bar h})=0,    \\
&&(h\bar h_0 +e_0\bar f +h_0\bar h + mf)(4-\frac{b\chi^2}{\bar
h})=0. \nonumber
\end{eqnarray}
Since we require $\chi^2-\frac{4\bar h}{b}<0$ for $\rho$ to be
complex, it turns out that
\begin{equation}
hh_{0} + e_{0}f=0,~~~ \bar{h}_{0}\bar h+m\bar f=0,~~~ h\bar h_0
+e_0\bar f +h_0\bar h + mf=0. \label{PQHF-Result}
\end{equation}
From Eqs.~(\ref{D7tad1})-(\ref{D3tad}), the D7, NS7, I7, and D3
flux contributions to the corresponding tadpoles are
\begin{eqnarray}
N_{flux\rm D7}:&&\frac{1}{2}[mh-e_0\bar h-(2b+\beta)q-(2\bar
b+\bar\beta)e] =\frac{1}{6}(mh -e_0\bar h)(4-\frac{b\chi^2}{\bar h}), \label{d7}\\
N_{flux\rm NS7}:&&\frac{1}{2}[\bar f h_0-f\bar h_0-\bar
a(2g+\gamma)+a(2\bar g+\bar \gamma)]=\frac{1}{6}(\bar fh_0-f\bar h_0)(4-\frac{b\chi^2}{\bar h}), \label{ns7}\\
N_{flux\rm I7}:&& \frac{1}{2}[\bar h_0h -\bar hh_0 +\bar
a(2b+\beta) -a(2\bar b+\bar\beta) + e_0\bar f-mf +q(2g+\gamma)
+e(2\bar g+\bar \gamma)] \nonumber \\&&=\frac{1}{6}(h\bar h_0-\bar
h h_0
+e_0\bar f -mf)(4-\frac{b\chi^2}{\bar h}),  \label{i7}\\
N_{\rm D3}:&&16-\frac{1}{2}[mh_0-e_0\bar h_0+3qa +3e\bar a]
=16-\frac{1}{6}(mh_0 -e_0\bar h_0)(4-\frac{b\chi^2}{\bar h}).
\label{d3}
\end{eqnarray}

The third antisymmetry condition in Eq.~(\ref{antis3}) has not
been confined, and we will consider it in two cases, $(p,q)=(0,0)$
and $p\cdot q \neq 0$.

Here, we emphasize again that $N_{flux\rm D7}$, $N_{flux\rm NS7}$,
$N_{flux\rm I7}$, and $N_{\rm D3}$ are respectively the numbers of
D7-branes, NS7-branes, I7-branes, D3-branes that we need to
introduce for their tadpole cancellations due to the flux
contributions. In other words, $-N_{flux\rm D7}$, $-N_{flux\rm
NS7}$, $-N_{flux\rm I7}$, and $-N_{\rm D3}$ are the flux
contributions to the D7-brane tadpoles, NS7-brane tadpoles,
I7-brane tadpoles, and D3-brane plus O3-plane tadpoles,
respectively.

\subsubsection{$(p,q)=(0,0)$}

The seven-brane tadpole contributions of the fluxes must be zero
because we set $(p,q)=(0,0)$ in this analysis. From
Eqs.~(\ref{d7})-(\ref{i7}), we can conclude, for $b\neq 0$,
\begin{equation}
mh -e_0\bar h=0,~~~ \bar fh_0-f\bar h_0=0,~~~ h\bar h_0-\bar h h_0
+e_0\bar f -mf=0.
\end{equation}
From the tadpole condition of the NS7-branes it results in zero
imaginary part of $\tau$. Thus, the moduli $\tau_i$ can not be
stabilized properly.

\subsubsection{$p\cdot q\neq 0$}

Recall the third antisymmetry condition in Eq.~(\ref{antis3}) for
$k\not =a$, $i=1,2,3$, we have
\begin{eqnarray}
\frac{m^I_{a}}{n^I_a}(f_ap^I_k-h_aq^I_k) &=&0  ,
\label{fqQ1}\\
\frac{m^I_{a}}{n^I_a}(g_{ia}p^I_k-b_{ia}q^I_k) &=&0  ,
\label{fqQ2}\\
\frac{m^I_{a}}{n^I_a}(\bar{g}_{ia}p^I_k-\bar{b}_{ia}q^I_k)&=&0,
\label{fqQ3}\\
\frac{m^I_{a}}{n^I_a}(\bar{f}_ap^I_k-\bar{h}_aq^I_k)&=&0.
\end{eqnarray}
From Eqs.~(\ref{xichi}) and (\ref{xi'chi'}) it is required that
$\xi=\xi'$ if $b,g,\bar h, \bar f, \xi, \xi'$ are non-zero. It
turns out that the imaginary part of $\sigma$ is zero. Thus, the
real part of dilaton $S$ can not be stabilized properly. If we can
tolerate Im$(\sigma)=0$, we can continue this analysis to see if
there is a solution for the remaining conditions. Let
$p^I_i=\mathbf{p}$, $q^I_i=\mathbf{q}$ for simplicity, the third
antisymmetry condition in Eq.~(\ref{antis3}) implies
\begin{equation}
\frac{f}{h}=\frac{\bar f}{\bar h}
=\frac{g}{b}=\frac{\gamma}{\beta} =\frac{\bar g}{\bar b}
=\frac{\bar\gamma}{\bar\beta} =\frac{\mathbf{q}}{\mathbf{p}}.
\end{equation}

We have known $b\bar h>0$ for $\rho_0$ complex, and this implies
$g\bar f>0$. Therefore for $\tau$ condition in Eq.~(\ref{tau-pq}),
it is required that $\bar f h_0-f\bar h_0<0$. However, this makes
the flux NS7-brane tadpole contribution Eq.~(\ref{ns7}) always
negative, {\it i.e.}, the flux contribution to the NS7-brane
tadpole is positive. Thus, we can not obtain supersymmetric
Minkowski flux vacua where the NS7-brane tadpole cancellation
condition is relaxed and the moduli are stabilized properly. For
the $(p,q)=(1,0)$ case, the discussions are similar.

\subsubsection{A Special Case with $b=0$}

Considering the case $(b,\bar h)=(0,0)$,  we obtain the following
conditions from the antisymmetry conditions and Jocobi identities:
\begin{eqnarray}
&&\beta=0, ~~~\bar\beta=-\bar b,~~~\gamma=-g, ~~~\bar\gamma=-\bar g; \nonumber \\
&&\bar b f=\bar g h, ~~~\bar b g=\bar f h,  ~~~f\bar f = g\bar g.
\end{eqnarray}
We can rewrite $E_3$ and $E_4$ as
\begin{eqnarray}
&&E_3=-3(h+\bar b\rho^2)=-\frac{3h}{f}(f+\bar g\rho^2), \\
&&E_4=3(f-g\rho+\bar g \rho^2-\bar f \rho^3)=-\frac{3}{f}(g\rho-f)
(f+\bar g\rho^2).
\end{eqnarray}
$\rho$ can have a pure imaginary root if $f\bar g>0$. The moduli
$\sigma$ and $\tau$ are
\begin{eqnarray}
&& \sigma = -\frac{E_3}{E_4} = -\frac{h}{g\rho-f},\\
&& \tau=-\frac{E_2}{E_4}= \frac{f}{3}\frac{h_0+3a\rho-3\bar a
\rho^2 -\bar h_0\rho^3}{(g\rho-f)(f+\bar g\rho^2)}.
\end{eqnarray}
For a finite $\tau$, $E_2$ also has the factor $(f+\bar g\rho^2)$,
so we obtain two additional constraints
\begin{eqnarray}
h_0\bar g+3\bar a f=0, ~~~\bar h_0 f +3a\bar g=0.
\end{eqnarray}
The conditions for Im$(\sigma)$ and Im$(\tau)$ are
\begin{eqnarray}
{\rm Im}(\sigma):~ h\bar f>0,~~~ {\rm Im}(\tau):~h_0\bar f+3a \bar
g<0.
\end{eqnarray}

On the other hand, $E$ has double roots and can be rewritten in
terms of $E_3$ and $E_4$ as
\begin{eqnarray}
E = \frac{3}{f}(f+\bar g\rho^2)(hE_2-(g\rho-f)E_1) =
\frac{3}{f}(f+\bar g\rho^2)^2(\mathcal{A}\rho^2 +\mathcal{B}\rho
+\mathcal{C} ).
\end{eqnarray}
where the coefficients $\mathcal{A}$, $\mathcal{B}$, and
$\mathcal{C}$ are
\begin{eqnarray}
\mathcal{A}=\frac{mg}{\bar g},~~~\mathcal{B}=-\frac{1}{\bar
g}(h\bar h_0+mf+3gq),~~~\mathcal{C}=e_0+\frac{hh_0}{f}.
\end{eqnarray}
Combining with the antisymmetry constraints in
Eqs.~(\ref{qhfpN1}), (\ref{qhfpN2}), and (\ref{qhfpN4}), we obtain
the following relations:
\begin{eqnarray}
&& e_0\bar g +\bar b h_0 -\bar a h +fq=0, ~~~\bar h_0h +e_0\bar f -a\bar b+gq =0,\nonumber \\
&& e\bar f=mg, ~~~ e\bar g=mf,~~~ eg=fq-\bar a h, ~~~ e_0 \bar
f=2a\bar b+2e\bar g.
\end{eqnarray}

The seven-brane tadpole contributions of the fluxes are
\begin{eqnarray}
N_{flux \rm D7}:&& \frac{1}{2}[mh-e_0\bar h-(2b+\beta)q-(2\bar
b+\bar\beta)e] =\frac{1}{2}(mh -e\bar b)=0,  \\
N_{flux\rm NS7}:&&\frac{1}{2}[\bar f h_0-f\bar h_0-\bar
a(2g+\gamma)+a(2\bar g+\bar \gamma)]=\frac{2}{3}[3a\bar g+h_0\bar f],\\
N_{flux\rm I7}:&& \frac{1}{2}[\bar h_0h -\bar hh_0 +\bar
a(2b+\beta) -a(2\bar b+\bar\beta) + e_0\bar f-mf +q(2g+\gamma)
+e(2\bar g+\bar \gamma)] \nonumber \\&&=\frac{1}{2}[h\bar
h_0-a\bar b+e_0\bar f+qg]=0.
\end{eqnarray}

It is generic that $N_{flux\rm D7}=0$ and $N_{flux\rm I7}=0$ from
the conditions above. We consider two possible cases for the
NS7-brane tadpole in the following discussion:

\begin{itemize}

\item $(p,q)=(0,0)$

The condition $q^I=0$ implies $N_{flux\rm NS7}=0$, which turns out
$3a\bar g +h_0 \bar f = 0$. Then there is no imaginary part for
$\tau$. Thus, the moduli $\tau_i$ can not be stabilized properly.

\item $p\cdot q\neq 0$

A nonzero $q^I=0$ implies $N_{flux\rm NS7}>0$, however it violates
the condition of Im$(\tau)$ for it to have the same sign as
Im$(\rho)$.  Thus again, we do not have supersymmetric Minkowski
flux vacua where the NS7-brane tadpole cancellation condition can
be relaxed and the moduli can be stabilized properly.

\end{itemize}

\section{Supersymmetric $\bf AdS$ Vacua  and
A Semi-Realistic Pati-Salam Model}

To relax the constraints we shall consider an AdS vacuum. For
simplicity, we choose $q^I_i=0$ so that $P=0$ for the third
antisymmetry condition relaxed. In addition, we assume $E_2=0$ by
ignoring the dilaton modulus at the current stage because there
are enough degrees of freedom to compute it at any time. By the
AdS conditions, we obtain
\begin{equation}
E_3=0,~~~ \tau=-\frac{E_1'}{E_3'}.
\end{equation}
Again with the same setup, we have
\begin{equation}
h =b{\xi},~~~ b+\beta=h\chi=b\xi\chi,~~~ \bar b =\bar h\xi,~~~
\bar b+\bar\beta=b\chi~,~\,
\end{equation}
and $E_3$ is factorized as $
E_3=3(\rho-\xi)(\bar h \rho^2-b\chi\rho +b)$. The complex roots of
$\rho$ have the following properties
\begin{equation}
\rho_0=\frac{bt\pm\sqrt{b^2\chi^2-4b\bar h}}{2\bar
h},~~~b^2\chi^2-4b\bar h<0,~~~ b^2\chi^2<4b\bar h,~~~ b\bar h>0~.~\,
\end{equation}
The modulus $\tau_0$ and the condition for its imaginary part to
have the same sign as $\rho_0$ are
\begin{eqnarray}
&&\tau_0=-\frac{E_1'}{E_3'}|_{\rho_0}=
-\frac{3e+6q\rho_0-3m\rho_0^2}{3(\rho_0-\xi)(2\bar
h\rho_0-b\chi)}, \\
&&\frac{1}{\bar h}(mb^2\xi\chi^2-mb^2\chi-2qb\bar h \xi\chi
+4qb\bar h -2mb\bar h\xi +eb\bar h \chi-2e\bar h^2 \xi)>0.
\label{tauAds}
\end{eqnarray}

The D7-brane tadpole contribution of the fluxes is
\begin{eqnarray}
N_{flux \rm D7}&=&\frac{1}{2}[mh-e_0\bar h-(2b+\beta)q-(2\bar
b+\bar\beta)e] \\
&=&\frac{1}{2}[mb\xi-e_0\bar h-qb-qb\xi\chi-eb\chi-e\bar h \xi]~.~\,
\end{eqnarray}

Therefore, for the case of $N_{flux \rm D7}>0$, Im$(\tau_0)$ can
still be positive, which implies the existence of AdS solutions.
The flux contributions to the D3-brane,  NS7-brane, and I7-brane
tadpoles are
\begin{eqnarray}
N_{flux \rm D3}:&& 16-\frac{1}{2}[mh_0-e_0\bar h_0+3qa+3e\bar a]
=16=\sum_In^I_in^I_jn^I_k,\\
N_{flux \rm NS7}:&&\frac{1}{2}[\bar f h_0-f\bar h_0-\bar
a(2g+\gamma)+a(2\bar g+\bar \gamma)]=0,\\
N_{flux \rm I7}:&& \frac{1}{2}[\bar h_0h -\bar hh_0 +\bar a(2b+\beta)
-a(2\bar b+\bar\beta) + e_0\bar f-mf +q(2g+\gamma) +e(2\bar g+\bar
\gamma)] \nonumber \\&&=0.
\end{eqnarray}


\begin{table}[h]
\begin{center}
\begin{tabular}{|c|c|c|c|c|c|c|c|c|c|c|c|} \hline

$b_1$ & $b_2$ & $b_3$ &  $m$ & $e_0$ & $e$ & $q$ \\ \hline

$4$ & $12$ & $12$ & $-2$ & $2$ & $-2$ & $-2$  \\
\hline

\end{tabular}
\caption{The choices of fluxes.} \label{Fluxes}
\end{center}
\end{table}

Next, let us construct a semi-realistic Pati-Salam model. We
assume $\chi=1$, $\xi=1$, $b=h=\bar h =\bar b$, and
$\beta=\bar\beta=0$. From Eq.~(\ref{tauAds}), for Im$(\tau)$
having the same sign as Im$(\rho)$, we obtain
\begin{equation}
b(2q-2m-e)>0.
\end{equation}
The D7-brane tadpole from flux contribution is
\begin{equation}
N_{flux\rm D7}=
\frac{1}{2}b(m-e_0-2q-2e)=\frac{1}{2}[-b(2q-2m-e)-b(m+e_0+3e)]>0.
\end{equation}
We assume $E_2=0$ by setting $a=\bar a=h_0=\bar h_0=0$, so then
$N_{\rm D3}=16$. We shall present an example by choosing $N_{\rm
D7}=8$ and $(e_0+m+3e)\neq0$. Recall that we assumed $\tau b\equiv
(\tau_i/k_i) (b_i k_i)$, with the Freed-Witten anomaly condition
imposed, we obtain the choices of the fluxes which are given in
Table~\ref{Fluxes}. We present the D-brane configurations and
intersection numbers of our Pati-Salam model in
Table~\ref{D-branes}. The corresponding particle spectrum is given
in Table~\ref{Spectrum}.

\begin{table}[h]
\begin{center}
\footnotesize
\renewcommand{\arraystretch}{1}
\begin{tabular}{|c|c|ccc|c|c|c|c|c|c|c|c|c|c|}
\hline

stack & $N$ & ($n_1$,$l_1$) & ($n_2$,$l_2$) & ($n_3$,$l_3$) & A &
S & $b$ & $b'$ & $c$ & $c'$ & $d$ & $d'$ & $e$ & $f$   \\
\hline \hline

$a$ & 4 & ( 2, 0) & ( 1,-1) & ( 1, 1) & 0(-1) & 0 & 3 & 0(3) & -3
& 0(3) & 0(-2) & 0(2) & 2 & -2  \\ \hline

$b$ & 2 & ( 1,-3) & ( 1, 1) & ( 2, 0) & 0(-3) & 0 & - & - & 3 &
0(1) & 1 & 0(1) & 0(1) & 2   \\  \hline

$c$ & 2 & ( 1, 3) & ( 2, 0) & ( 1,-1) & 0(-3) & 0 & - & - & - & -
& 0(-1) & 2 & -2 & 0(-1) \\ \hline \hline

$d$ & 2 & ( 0, 2) & ( 1,-1) & ( 1,-1) & -1 & -1 & - & - & - & - &
- & - & 0(1) & 0(1)  \\ \hline

$e$ & 1 & ( 0, 2) & ( 0,-2) & ( 2, 0) & - & - & - & - & - & - & -
& - & - & 0(-4) \\ \hline

$f$ & 1 & ( 0, 2) & ( 2, 0) & ( 0,-2) & - & - &
\multicolumn{8}{|c|}{$3\chi_1=\chi_2=\chi_3$} \\ \hline

\end{tabular}
\caption{D-brane configurations and intersection numbers for a
model on $\mathbf{T}^6$ orientifold. The complete gauge symmetry
is $[U(4)_C \times U(2)_L \times U(2)_R]_{observable}\times
[U(2)\times USp(2)^2]_{hidden}$, the SM fermions and Higgs fields
arise from the first two torus. }
\label{D-branes}
\end{center}
\end{table}

\begin{table}[h]
\renewcommand{\arraystretch}{.75}
\caption{The chiral and vector-like superfields, and their quantum
numbers under the gauge symmetry $U(4)_C\times U(2)_L\times U(2)_R
\times U(2)\times USp(2)^2$.} \label{Spectrum}
\begin{center}
\begin{tabular}{|c||c||c|c|c||c|c|c|}\hline
 & Quantum Number
& $Q_4$ & $Q_{2L}$ & $Q_{2R}$  & Field \\
\hline\hline
$ab$ & $4 \times (4,\bar{2},1,1,1,1)$ & 1 & -1 & 0  & $F_L(Q_L, L_L)$\\
$ac$ & $4\times (\bar{4},1,2,1,1,1)$ & -1 & 0 & $1$   & $F_R(Q_R, L_R)$\\
\hline\hline
$ae$ & $2 \times (4,1,1,1,\bar 2,1)$ & 1 & 0 & 0  & \\
$af$ & $2 \times (\bar 4,1,1,1,1, \bar 2)$ & -1 & 0 & 0  &      \\
\hline
$ac'$ & $3 \times (4,1,2,1,1,1)$ & 1 & 0 & 1  & \\
& $3 \times (\bar{4},1,\bar{2},1,1,1)$ & -1 & 0 & -1  & \\
$ad$ & $2 \times (4,1,1,\bar 2,1,1)$ & 1 & 0 & 0  & \\
& $2 \times (\bar{4},1,1,2,1,1)$ & -1 & 0& 0  & \\
$ad'$ & $2 \times (4,1,1,2,1,1)$ & 1 & 0 & 0  & \\
& $2 \times (\bar{4},1,1,\bar{2},1,1)$ & -1 & 0 & 0  & \\
\hline
$bd$ & $1 \times (1,2,1,\bar 2,1,1)$ & 0 & 1 & 0  & \\
$bf$ & $2 \times (1,2,1,1,1,\bar 2)$ & 0 & 1 & 0  & \\
\hline
$bc'$ & $1 \times (1,2,2,1,1,1)$ & 0 & 1 & 1  & \\
& $1 \times (1,\bar{2},\bar{2},1,1,1)$ & 0 & -1 & -1  & \\
$bd'$ & $1 \times (1,2,1,2,1,1)$ & 0 & 1 & 0  & \\
& $1 \times (1,\bar{2},1,\bar 2,1,1)$ & 0 & -1 & 0  & \\
$be$ & $1 \times (1,2,1,1,\bar 2,1)$ & 0 & 1 & 0  &  \\
& $1 \times (1,\bar{2},1,1,2,1)$ & 0 & -1 & 0 &\\
\hline
$cd'$ & $2 \times (1,1,2,2,1,1)$ & 0 & 0 & 1  &  \\
$ce$ & $2 \times (1,1,\bar 2,1,2,1)$ & 0 & 0 & -1  &  \\
\hline
$cd$ & $1 \times (1,1,2,\bar{2},1,1)$ & 0 & 0 & 1  &  \\
 & $1 \times (1,1,\bar{2},2,1,1)$ & 0 & 0 & -1 &  \\
$cf$ & $1 \times (1,1,2,1,1,\bar{2})$ & 0 & 0 & 1  &  \\
 & $1 \times (1,1,\bar{2},1,1,2)$ & 0 & 0 & -1 &  \\
\hline
$bc$ & $3 \times (1,2,\bar 2,1,1,1)$ & 0 & 1 & -1  & $H_u^i$, $H_d^i$\\
\hline
\end{tabular}
\end{center}
\end{table}




\section{Discussions and Conclusion}

We conjectured that in generic supsersymmetric Minkowski vacua, at
least one of the flux contributions to the seven-brane and
D3-brane tadpoles are positive if the moduli can be stabilized
properly on the Type IIB toroidal $\mathbf{T^6}$ orientifolds with
the RR, NSNS, metric, non-geometric and S-dual flux
compactifications. Therefore, these tadpole cancellation conditions
can not be relaxed for realistic model building. To study the
supsersymmetric Minkowski vacua, we started from the reasonably
simplified fluxes and then discussed the corresponding
superpotential. We showed that we are not able to have the
positive real parts of all the moduli and the negative/zero flux
contributions to all the seven-brane and D3-brane tadpoles
simultaneously. In the supsersymmetric AdS vacua, we can have the
flux vacua where the seven-brane and D3-brane tadpole cancellation
conditions are relaxed elegantly, and we presented a concrete
semi-realistic Pati-Salam model as well as its particle spectrum.
The lifting from the AdS vacua to the Minkowski/dS vacua remains a
great challenge in flux model building and is dedicated to the
future work. On the other hand, some directions of searching for
Minkowsky/dS vacua with fluxes on general geometries are
interesting, for example, the recent development with structure
manifolds \cite{Danielsson:2011au}.

\section*{Acknowledgments}

We would like to thank Y.~Liu very much
for the collaborations in the early
stage of this project, and thank P.~G.~Camara for helpful discussions.
This research was supported in part by the Austrian FWF project
P21239 (CMC), by the Mitchell-Heep Chair in High Energy Physics
(SH), by the DOE grant DE-FG03-95-Er-40917 (TL and DVN),
and by the Natural Science Foundation of China
under grant numbers 10821504 and 11075194 (TL).


\begin{thebibliography}{99}
\itemsep 0.5mm





\bibitem{Polchinski:1995df}
  J.~Polchinski and E.~Witten,
  Nucl.\ Phys.\  B {\bf 460}, 525 (1996).

\bibitem{Berkooz:1996km}
  M.~Berkooz, M.~R.~Douglas and R.~G.~Leigh,
  Nucl.\ Phys.\  B {\bf 480}, 265 (1996).

\bibitem{Bachas:1995ik}
  C.~Bachas,
  arXiv:hep-th/9503030.


\bibitem{Blumenhagen:2005mu}
  R.~Blumenhagen, M.~Cveti\v c, P.~Langacker and G.~Shiu,
  Ann.\ Rev.\ Nucl.\ Part.\ Sci.\  {\bf 55}, 71 (2005),
and the references therein.



\bibitem{Blumenhagen:2000wh}
  R.~Blumenhagen, L.~Goerlich, B.~Kors and D.~Lust,
  JHEP {\bf 0010}, 006 (2000).

\bibitem{Angelantonj:2000hi}
  C.~Angelantonj, I.~Antoniadis, E.~Dudas and A.~Sagnotti,
  Phys.\ Lett.\ B {\bf 489}, 223 (2000).



\bibitem{Ibanez:2001nd}
  L.~E.~Ibanez, F.~Marchesano and R.~Rabadan,
  JHEP {\bf 0111}, 002 (2001).








\bibitem{CSU1}
M.~Cveti\v c, G.~Shiu and A.~M.~Uranga, Phys.\ Rev.\ Lett.\  {\bf
87}, 201801 (2001).

\bibitem{CSU2}
M.~Cveti\v c, G.~Shiu and A.~M.~Uranga, Nucl.\ Phys.\ B {\bf 615},
3 (2001).

\bibitem{CP} M. Cveti\v c and I. Papadimitriou,
Phys.\ Rev.\ D {\bf 67}, 126006 (2003).

\bibitem{Cvetic:2002pj}
  M.~Cveti\v c, I.~Papadimitriou and G.~Shiu,
  Nucl.\ Phys.\ B {\bf 659}, 193 (2003)
  [Erratum-ibid.\ B {\bf 696}, 298 (2004)].

\bibitem{CLL}
M.~Cveti\v c, T.~Li and T.~Liu,
Nucl.\ Phys.\ B {\bf 698}, 163 (2004).

\bibitem{Cvetic:2004nk}
  M.~Cveti\v c, P.~Langacker, T.~Li and T.~Liu,
  Nucl.\ Phys.\ B {\bf 709}, 241 (2005).


\bibitem{Chen:2005ab}
  C.-M.~Chen, G.~V.~Kraniotis, V.~E.~Mayes, D.~V.~Nanopoulos and J.~W.~Walker,
  Phys.\ Lett.\ B {\bf 611}, 156 (2005);
  Phys.\ Lett.\  B {\bf 625}, 96 (2005).


\bibitem{Chen:2005mj}
  C.~M.~Chen, T.~Li and D.~V.~Nanopoulos,
  Nucl.\ Phys.\ B {\bf 732}, 224 (2006).



\bibitem{Chen:2006sd}
  C.~M.~Chen, V.~E.~Mayes and D.~V.~Nanopoulos,
  Phys.\ Lett.\  B {\bf 648}, 301 (2007).





\bibitem{ListSUSYOthers}
R.~Blumenhagen, L.~G\"orlich and T.~Ott, JHEP {\bf 0301}, 021 (2003);
G.~Honecker, Nucl.\ Phys.\  {\bf B666}, 175 (2003);
G.~Honecker and T.~Ott,
Phys.\ Rev.\ D {\bf 70}, 126010 (2004)
  [Erratum-ibid.\ D {\bf 71}, 069902 (2005)].


\bibitem{Chen:2007px}
  C.~-M.~Chen, T.~Li, V.~E.~Mayes, D.~V.~Nanopoulos,
  Phys.\ Lett.\  {\bf B665}, 267-270 (2008).

\bibitem{Chen:2007zu}
  C.~-M.~Chen, T.~Li, V.~E.~Mayes, D.~V.~Nanopoulos,
  Phys.\ Rev.\  {\bf D77}, 125023 (2008).





\bibitem{CLW}
M.~Cveti\v c, P.~Langacker and J.~Wang,
Phys.\ Rev.\ D {\bf 68}, 046002 (2003).


\bibitem{GVW}
S.~Gukov, C.~Vafa and E.~Witten,
Nucl.\ Phys.\ B {\bf 584}, 69 (2000) [Erratum-ibid.\ B {\bf 608}, 477 (2001)].




\bibitem{Giddings:2001yu}
  S.~B.~Giddings, S.~Kachru and J.~Polchinski,
  Phys.\ Rev.\  D {\bf 66}, 106006 (2002).

\bibitem{Kachru:2002sk}
  S.~Kachru, M.~B.~Schulz, P.~K.~Tripathy and S.~P.~Trivedi,
  JHEP {\bf 0303}, 061 (2003).



\bibitem{Kachru:2003aw}
  S.~Kachru, R.~Kallosh, A.~Linde and S.~P.~Trivedi,
  Phys.\ Rev.\  D {\bf 68}, 046005 (2003).





\bibitem{CU}
J. F. G. Cascales and A. M. Uranga,
JHEP {\bf 0305}, 011 (2003).

\bibitem{BLT}
R.~Blumenhagen, D.~L\"ust and T. R. Taylor,
Nucl.\ Phys.\ B {\bf 663}, 319 (2003).




\bibitem{MS}
F.~Marchesano and G.~Shiu,
Phys.\ Rev.\ D {\bf 71}, 011701 (2005);
JHEP {\bf 0411}, 041 (2004).

\bibitem{CL}
M.~Cveti\v c and T.~Liu, Phys.\ Lett.\ B {\bf 610}, 122 (2005).

\bibitem{Cvetic:2005bn}
  M.~Cveti\v c, T.~Li and T.~Liu,
  Phys.\ Rev.\ D {\bf 71}, 106008 (2005).


\bibitem{Kumar:2005hf}
  J.~Kumar and J.~D.~Wells,
  JHEP {\bf 0509}, 067 (2005).



\bibitem{Chen:2005cf}
  C.~M.~Chen, V.~E.~Mayes and D.~V.~Nanopoulos,
  Phys.\ Lett.\  B {\bf 633}, 618 (2006).





\bibitem{Grimm:2004ua}
  T.~W.~Grimm and J.~Louis,
  Nucl.\ Phys.\ B {\bf 718}, 153 (2005).


\bibitem{Villadoro:2005cu}
  G.~Villadoro and F.~Zwirner,
  JHEP {\bf 0506}, 047 (2005).


\bibitem{Camara:2005dc}
  P.~G.~Camara, A.~Font and L.~E.~Ibanez,
  JHEP {\bf 0509}, 013 (2005).



\bibitem{Chen:2006gd}
  C.~M.~Chen, T.~Li and D.~V.~Nanopoulos,
  Nucl.\ Phys.\  B {\bf 740}, 79 (2006).


\bibitem{Chen:2006ip}
  C.~M.~Chen, T.~Li and D.~V.~Nanopoulos,
  Nucl.\ Phys.\  B {\bf 751}, 260 (2006).



\bibitem{Shelton:2005cf}
  J.~Shelton, W.~Taylor and B.~Wecht,
  JHEP {\bf 0510}, 085 (2005).

\bibitem{Aldazabal:2006up}
  G.~Aldazabal, P.~G.~Camara, A.~Font and L.~E.~Ibanez,
  JHEP {\bf 0605}, 070 (2006).


\bibitem{Villadoro:2006ia}
  G.~Villadoro and F.~Zwirner,
  JHEP {\bf 0603}, 087 (2006).




\bibitem{Chen:2007af}
  C.~-M.~Chen, T.~Li, Y.~Liu, D.~V.~Nanopoulos,
  Phys.\ Lett.\  {\bf B668}, 63-66 (2008).



\bibitem{Freed:1999vc}
  D.~S.~Freed and E.~Witten,
  arXiv:hep-th/9907189.



\bibitem{Cascales:2003zp}
  J.~F.~G.~Cascales and A.~M.~Uranga,
  JHEP {\bf 0305}, 011 (2003).

\bibitem{Chen:2008ht}
  C.~M.~Chen, T.~Li and D.~V.~Nanopoulos,
  arXiv:0812.2089 [hep-th].

\bibitem{Aldazabal:2008zza}
  G.~Aldazabal, P.~G.~Camara, J.~A.~Rosabal,
  Nucl.\ Phys.\  {\bf B814}, 21-52 (2009).

\bibitem{deCarlos:2009fq}
  B.~de Carlos, A.~Guarino and J.~M.~Moreno,
  JHEP {\bf 1001}, 012 (2010);
  JHEP {\bf 1002}, 076 (2010).

\bibitem{Cremades:2002te}
  D.~Cremades, L.~E.~Ibanez, F.~Marchesano,
  JHEP {\bf 0207}, 009 (2002).

\bibitem{Lust:2004cx}
  D.~Lust, P.~Mayr, R.~Richter, S.~Stieberger,
  Nucl.\ Phys.\  {\bf B696}, 205-250 (2004).


\bibitem{Aldazabal:2000dg}
  G.~Aldazabal, S.~Franco, L.~E.~Ibanez, R.~Rabadan and A.~M.~Uranga,
  J.\ Math.\ Phys.\  {\bf 42}, 3103 (2001).

\bibitem{Danielsson:2011au}
  U.~H.~Danielsson, S.~S.~Haque, P.~Koerber, G.~Shiu, T.~Van Riet and T.~Wrase,
  arXiv:1103.4858 [hep-th].


\end{thebibliography}
\end{document}